\documentclass{PoS}
\usepackage{graphicx}
\usepackage{amsmath}
\usepackage{psfrag}
\usepackage{slashed}
\usepackage[sort&compress,numbers]{natbib}
\def\mc#1{\mathcal{#1}}
\def\tr{\mathop{\rm tr}\nolimits}

\def\phid{\phi^\dagger}

\def\I33m{\mathrm{I}_3^{3{\mathrm m}}}
\def\nn{\nonumber}
\def\flo{\rm{f}}
\def\be{\begin{equation}}
\def\ee{\end{equation}}
\def\bea{\begin{eqnarray}}
\def\eea{\end{eqnarray}}
\def\treenum{{(0)}}
\def\oneloopnum{{(1)}}

\def\qb{{\bar{q}}}
\def\cg{c_{\Gamma}}
\def\e{\epsilon}

\def\spa#1.#2{\left\langle#1\,#2\right\rangle}
\def\spb#1.#2{\left[#1\,#2\right]}
\def\lor#1.#2{\left(#1\,#2\right)}
\def\sand#1.#2.#3{%
\left\langle\smash{#1}{\vphantom1}^{-}\right|{#2}%
\left|\smash{#3}{\vphantom1}^{-}\right\rangle}
\def\sandp#1.#2.#3{%
\left\langle\smash{#1}{\vphantom1}^{-}\right|{#2}%
\left|\smash{#3}{\vphantom1}^{+}\right\rangle}
\def\sandpp#1.#2.#3{%
\left\langle\smash{#1}{\vphantom1}^{+}\right|{#2}%
\left|\smash{#3}{\vphantom1}^{+}\right\rangle}
\def\sandpm#1.#2.#3{%
\left\langle\smash{#1}{\vphantom1}^{+}\right|{#2}%
\left|\smash{#3}{\vphantom1}^{-}\right\rangle}
\def\sandmp#1.#2.#3{%
\left\langle\smash{#1}{\vphantom1}^{-}\right|{#2}%
\left|\smash{#3}{\vphantom1}^{+}\right\rangle}
\def\spab#1.#2.#3{\langle#1|#2|#3]}
\def\spba#1.#2.#3{[#1|#2|#3\rangle}

\def\spa#1.#2{\left\langle#1\,#2\right\rangle}
\def\spb#1.#2{\left[#1\,#2\right]}
\def\spab#1.#2.#3{\left\langle#1\,#2\,#3\right]}

\def\spa#1.#2{\left\langle#1\,#2\right\rangle}
\def\spb#1.#2{\left[#1\,#2\right]}
\def\spab#1.#2.#3{\left\langle#1\,#2\,#3\right]}
\def\spba#1.#2.#3{\left[#1\,#2\,#3\right\rangle}
\def\spaa#1.#2.#3.#4.#5.#6{\left\langle#1\,#2\,#3\,#4\.#5\,#6\right\rangle}
\def\spbb#1.#2.#3.#4.#5.#6{\left [#1\,#2\,#3\,#4\,#5\,#6\right ]}
\def\spbnin#1.#2.#3.#4.#5.#6.#7.#8.#9{\left [#1|\,#2\,#3\,#4\,#5\,#6\,#7\,#8|\,#9\right \rangle}
\def\spanin#1.#2.#3.#4.#5.#6.#7.#8.#9{\left \langle#1|\,#2\,#3\,#4\,#5\,#6\,#7\,#8|\,#9\right]}
\def\spasev#1.#2.#3.#4.#5.#6.#7{\left \langle#1|\,#2\,#3\,#4\,#5\,#6\,|#7\right]}
\def\spafiv#1.#2.#3.#4.#5{\left \langle#1|\,#2\,#3\,#4\,|#5\right]}
\def\spbsev#1.#2.#3.#4.#5.#6.#7{\left [#1|\,#2\,#3\,#4\,#5\,#6|\,#7\right \rangle}
\def\spbsix#1.#2.#3.#4.#5.#6{\left [#1|\,#2\,#3\,#4\,#5|\,#6\right ]}
\def\spbfiv#1.#2.#3.#4.#5{\left [#1|\,#2\,#3\,#4|\,#5\right \rangle}
\def\spbf#1.#2.#3.#4{\left [#1\,#2\,#3\,#4\right ]}
\def\spahr#1.#2{\langle#1\,\hat{#2}\rangle}
\def\spaah#1.#2.#3.#4{\langle#1\,#2\,#3\,\hat{#4}\rangle}
\def\spaahl#1.#2.#3.#4{\langle\hat{#1}\,#2\,#3\,#4\rangle}
\def\spabh#1.#2.#3{\langle#1\,\widehat{#2}\,#3]}
\def\spahl#1.#2{\langle\hat{#1}\,#2\rangle}
\def\spahh#1.#2{\langle\hat{#1}\,\hat{#2}\rangle}
\def\spaas#1.#2.#3{\left\langle#1\,#2\,#3\right\rangle}
\def\spbhl#1.#2{\left[\hat{#1}\,#2\right]}
\def\spbhr#1.#2{\left[#1\,\hat{#2}\right]}
\def\spabhh#1.#2.#3{\langle\hat{#1}\,\hat{#2}\,#3]}
\def\spbah#1.#2.#3{[#1\,\widehat{#2}\,#3\rangle}
\def\mc#1{\mathcal{#1}}

\def\e{\epsilon}
\def\qb{{\bar{q}}}

\title{
\rule{0cm}{2.5cm}\vspace{-3.5cm}\\{ 
\it \normalsize \hspace{13cm} IPPP/09/98
} \vspace{2cm}\\Higgs $+ 2$ jets: Compact Analytic Results}

\ShortTitle{H$+2$ jets: Compact Analytic Results}

\author{\speaker{Ciaran Williams}\thanks{Based upon work done in collaboration with Simon Badger, John Campbell, Keith Ellis, Nigel Glover and Pierpaolo Mastrolia}\\
      Institute for Particle Physics Phenomenology  \\ Department of Physics \\University of Durham \\ Durham \\DH1 3LE \\ UK\\
        E-mail: \email{ciaran.williams@durham.ac.uk}}

\abstract{This report describes the recent efforts to compute analytic formulae for the Next-to-Leading-Order (NLO) QCD corrections to Higgs plus two jet production at hadron colliders. In these calculations the Higgs boson couples to gluons via a top-quark loop which is integrated out to form an effective vertex. The amplitudes are further simplified by splitting the real Higgs scalar into the sum of two complex scalars $\phi$ and $\phi^{\dagger}$. Four-dimensional unitarity is used to construct the cut-containing pieces of the amplitude, while a variety of bootstrap and Feynman diagram techniques are used to construct the rational pieces.  The results described here are valid in the limit of a large top quark mass and when the transverse momenta of the jets are less than $m_t$. }

\FullConference{RADCOR 2009 - 9th International Symposium on Radiative Corrections (Applications of Quantum Field Theory to Phenomenology) ,\\
		 October 25 - 30 2009\\
		 Ascona, Switzerland}

\begin{document}

\section{Introduction}

With the LHC now taking data and the Tevatron regularly setting new luminosity records, the search for the Higgs boson is entering a crucial phase. A discovery, or at the very least improved exclusion limits \cite{2009pt}, is to be expected in the near future. In order to maximise the effectiveness of the experimental search strategies it is crucial to have excellent theoretical predictions for both Higgs signal and background processes. 

One such process is the production of the Higgs boson in association with two jets. Here the signal provided by electroweak induced vector-boson-fusion (VBF) ~\cite{Djouadi:2005gi} is essential for a measurement of the Yukawa coupling of the W and Z vector bosons. The VBF process is under good theoretical control, with both strong and electroweak one-loop calculations completed. 

On the other hand one of the main background processes, Higgs boson production in association with two jets via gluon fusion has had less theoretical attention. One-loop QCD corrections to this process have been  calculated semi-numerically \cite{Ellis:2005qe,Campbell:2006xx}, which allowed some phenomenology to be done. Over the last couple of years much theoretical effort has been devoted to completing the analytic calculation of the Higgs plus two jet process, with the hope of producing faster code for phenomenological studies. 

Gluon fusion proceeds via a top-quark loop,  calculations which include the full top quark mass are difficult. However, by integrating out the top quark from the loop one can simplify the calculation. This procedure leads to an effective Lagrangian to express the coupling of gluons to 
the Higgs field~\cite{Wilczek:1977zn},
\be
\mathcal{L}_H^{\mathrm{int}} = \frac{C}{2} \, H\,\tr
G_{\mu\nu}\,G^{\mu\nu}\, . \label{Lint} \ee
The effective Lagrangian 
approximation is valid  in the limit $m_H < 2 m_t$. 
To $\mc{O}(\alpha_S^2)$ the coefficient $C$  
is given by~\cite{Djouadi:1991tka,Dawson:1990zj},
\be
C =\frac{\alpha_S}{6 \pi v} \Big( 1 +\frac{11}{4 \pi} \alpha_S\Big)
 + {\cal O}(\alpha_S^3) \;.
\ee
Here $v$ is the vacuum expectation value of the Higgs field (246 GeV). 
The trace in Eq.~(\ref{Lint}) is over the colour degrees of freedom
which, since SU(3) generators in the fundamental representation are normalised 
such that $\tr T^a T^b = \delta^{ab}$.
Introducing a complex scalar field~\cite{Dixon:2004za,Badger:HpartonMHV}, $
\phi = \frac{1}{2} \left( H+i A \right),\;\phid = \frac{1}{2} \left( H-iA \right) \;, 
$ results in the following expression for the effective Lagrangian, Eq.~(\ref{Lint}),
\begin{eqnarray}
\mathcal{L}_{H,A}^{\mathrm{int}} 
&=& \frac{C}{2} \Big [H\,\tr G_{\mu\nu}\,G^{\mu\nu}+i A\,\tr G_{\mu\nu}\,{}^*G^{\mu\nu}\Big] \nn =  C \Big [\phi\,\tr G_{\scriptscriptstyle{SD}\; \mu\nu}\,G^{\mu\nu}_{\scriptscriptstyle{SD}}
            +\phid\,\tr G_{\scriptscriptstyle{ASD}\; \mu\nu}\,G^{\mu\nu}_{\scriptscriptstyle{ASD}} \Big]  \;,
\end{eqnarray}
where the gluon field strength has been separated into a self-dual and an anti-self-dual component,
\be G_{\scriptscriptstyle{SD}}^{\mu\nu} = \frac{1}{2}
(G^{\mu\nu}+{}^*G^{\mu\nu})\, , \quad
G_{\scriptscriptstyle{ASD}}^{\mu\nu} = \frac{1}{2}
(G^{\mu\nu}-{}^*G^{\mu\nu})\, , \quad {}^*G^{\mu\nu} \equiv
\frac{i}{2} \e^{\mu\nu\rho\sigma} G_{\rho\sigma}\, .
\ee
Calculations performed in terms of the field $\phi$ are simpler 
than the calculations for the Higgs boson and, moreover, the amplitudes for $\phid$
can be obtained by parity.
In the final stage, the full Higgs boson amplitudes are then written as a combination of  $\phi$ and $\phid$ components:
\bea
A(H,\{p_k\})&=&A(\phi,\{p_k\})+A(\phid,\{p_k\}) \;, 
\eea

\section{Method} 

One-loop amplitudes contain two pieces referred to as the cut-constructible and rational pieces. 
In massless gauge theories the cut-constructible part of one-loop amplitudes can be written as a sum over constituent basis integrals,
\bea
C_4(\phi,1^{\lambda_1},2^{\lambda_2},3^{\lambda_3},4^{\lambda_4})=\sum_{i}{C}_{4;i}{I}_{4;i} + \sum_{i}{C}_{3;i}{I}_{3;i} +\sum_{i}{C}_{2;i}{I}_{2;i}.
\eea
Here ${I}_{j;i}$ represents a $j$-point scalar basis integral, with a coefficient
$C_{j;i}$. The sum over $i$ represents the sum over the partitions of the external
momenta over the $j$ legs of the basis integral. 

Multiple cuts isolate different integral functions and allow the
construction of a linear system of equations from which the coefficients can be extracted. When
considering quadruple cuts of one-loop amplitudes, one is forced to consider complex momenta in
order to fulfill the on-shell constraints~\cite{Britto:2004nc}. The four
on-shell constraints are sufficient to isolate each four-point (box) 
configuration by freezing the loop momentum, thereby allowing the determination of the corresponding coefficient
by a purely algebraic operation.
To isolate the coefficients of lower-point integrals, one needs to cut fewer than four lines.
In this case the loop momenta is no longer completely determined, but, according to the number of cuts, 
some of its components are free variables. 
In this case the computation of the three- (triangle) and two-point (bubble) coefficients can also be reduced 
to algebraic procedures by exploiting the singularity structure of amplitudes in the complex-plane \cite{Forde:intcoeffs}.
Alternatively one can extract the coefficients of bubble- and triangle-functions by employing 
the spinor-integration technique \cite{Britto:sqcd,Britto:ccqcd}.
This method has recently inspired a novel technique for evaluating the double-cut phase-space 
integrals {\it via} Stokes' Theorem applied to functions of 
two complex-conjugated variables \cite{Mastrolia:2009dr}. 

In addition to the cut-constructible pieces one-loop QCD amplitudes contain pieces which cannot be reconstructed by four dimensional cuts. As such additional techniques have been used to calculate the one-loop $\phi$ + 4 parton amplitudes. For the $\phi$-MHV, $\phi$-all minus, $\phi q\overline{q}$
-MHV and $\phi q\overline{q}Q\overline{Q}$ helicity amplitudes the unitarity bootstrap method was employed \cite{Berger:genhels}. In this approach one calculates the rational piece of the amplitudes from four-dimensional BCFW recursion relations \cite{Britto:proof}.  For the $\phi$-NMHV and $\phi q\overline{q}$-NMHV amplitudes the rational piece was obtained from the reduction of Feynman diagrams.

\section{Higgs plus four gluon amplitudes}

This section summarises the calculation of the $\phi$ plus four gluon amplitudes in the limit of a large top quark mass. 
The tree level amplitudes linking a $\phi$ with $n$ gluons 
can be decomposed into colour ordered amplitudes as~\cite{Dawson:Htomultijet},
\begin{align}
	{\cal A}^{(0)}_n(\phi,\{k_i,\lambda_i,a_i\}) = 
	i C g^{n-2}
	\sum_{\sigma \in S_n/Z_n}
	\tr(T^{a_{\sigma(1)}}\cdots T^{a_{\sigma(n)}})\,
	A^{(0)}_n(\phi,\sigma(1^{\lambda_1},..,n^{\lambda_n})).
	\label{TreeColorDecompositionQ}
\end{align}
Here $S_n/Z_n$ is the group of non-cyclic permutations on $n$
symbols, and $j^{\lambda_j}$ labels the momentum $p_j$ and helicity
$\lambda_j$ of the $j^{\rm th}$ gluon, which carries the adjoint
representation index $a_i$.  
The one-loop amplitudes
follow the same colour ordering as the pure QCD amplitudes \cite{Bern:1994zx},
\begin{align}
	\mc A^{(1)}_n(\phi,\{k_i,\lambda_i,a_i\}) &= i C g^{n}
	\sum_{c=1}^{[n/2]+1}\sum_{\sigma \in S_n/S_{n;c}} G_{n;c}(\sigma)
	A^{(1)}_n(\phi,\sigma(1^{\lambda_1},\ldots,n^{\lambda_n}))
	\label{eq:1lhtogcolour}
\end{align}
where  $	G_{n;1}(1) = N_c \tr( T^{a_1}\cdots T^{a_n} )$ and $
	 G_{n;c}(1) =   \tr( T^{a_1}\cdots T^{a_{c-1} } )
			\tr( T^{a_c}\cdots T^{a_n} ) c>2. $
The sub-leading terms can be computed by summing over various permutations of the leading colour
amplitudes~\cite{Bern:1994zx}. Table ~\ref{Hggggrefs} indicates the references in which the various helicity contributions to $Hgggg$ have first been calculated analytically. 
\TABLE{
\begin{tabular}{|l|l|l|}
\hline
$H$ amplitude & $\phi$ amplitude & $\phid$ amplitude\\
\hline
${\cal A}(H,1^+,2^+,3^+,4^+)$ &${\cal A}(\phi,1^+,2^+,3^+,4^+)$~\cite{Berger:2006sh}  &${\cal A}(\phid,1^+,2^+,3^+,4^+)$~\cite{Badger:2006us} \\
${\cal A}(H,1^-,2^+,3^+,4^+)$ &${\cal A}(\phi,1^-,2^+,3^+,4^+)$~\cite{Berger:2006sh}  &${\cal A}(\phid,1^-,2^+,3^+,4^+)$~\cite{Badger:2009hw}  \\
${\cal A}(H,1^-,2^-,3^+,4^+)$ &${\cal A}(\phi,1^-,2^-,3^+,4^+)$~\cite{Badger:2007si}  &${\cal A}(\phid,1^-,2^-,3^+,4^+)$~\cite{Badger:2007si}  \\
${\cal A}(H,1^-,2^+,3^-,4^+)$ &${\cal A}(\phi,1^-,2^+,3^-,4^+)$~\cite{Glover:2008ffa} &${\cal A}(\phid,1^-,2^+,3^-,4^+)$~\cite{Glover:2008ffa}  \\
\hline
\end{tabular} 
\caption{$\phi$ and $\phid$ amplitudes needed to construct a given one-loop $Hgggg$ amplitude, together with the references where they can be obtained.
In all cases the $\phid$ amplitudes are constructed from the $\phi$ amplitudes given in the reference using the parity operation.
Results for all helicity combinations are also written, in uniform notation, in ref.~\cite{Badger:2009hw}.
}
\label{Hggggrefs}
}

\section{Higgs plus four parton amplitudes}

The colour decomposition of the $H\bar{q}qgg$ amplitudes is exactly the same as for the
case $\bar{q}qgg$ which was written down in ref.~\cite{Bern:1994fz}. At tree-level 
there are two colour stripped amplitudes,
\be
{\cal A}_4^\treenum(\phi,1_\qb,2_q,3,4)
 = Cg^2 \, \sum_{\sigma\in S_2}\left(
T^{a_{\sigma(3)}}T^{a_{\sigma(4)}}\right)_{i_2}^{\,\,\,\bar{\imath}_1}
A_4^\treenum(\phi,1_\qb,2_q,\sigma(3),\sigma(4)) \,.
\label{qqggtreedecomp}
\ee
At one-loop level the colour decomposition is, 
\bea
{\cal A}_4^\oneloopnum(\phi,1_\qb,2_q,3,4) &=& Cg^4\, \cg
\bigg[ N_c\sum_{\sigma\in S_2}\left(
T^{a_{\sigma(3)}}T^{a_{\sigma(4)}}\right)_{i_2}^{\,\,\,\bar{\imath}_1}
A_{4;1}(\phi,1_\qb,2_q,\sigma(3),\sigma(4))
\nonumber\\ && \hskip1.3cm \null
+ \delta^{a_3 a_4} \, \delta_{i_2}^{\,\,\,\bar{\imath}_1}
A_{4;3}(\phi,1_\qb,2_q;3,4) \bigg] \,.
\label{qqggloopdecomp}
\eea
The colour stripped amplitudes $A_{4;1}$ and $A_{4;3}$ 
can further be decomposed into primitive amplitudes,
\bea A_{4;1}(\phi,1_\qb,2_q,3,4) &=& A_4^L(\phi,1_\qb,2_q,3,4) -
\frac{1}{N_c^2}A_4^R(\phi,1_\qb,2_q,3,4) +\frac{n_f}{N_c} \, A_4^{\flo}(\phi,1_\qb,2_q,3,4) \,,
\label{A41defn}
\eea
and,
\bea A_{4;3}(\phi,1_\qb,2_q;3,4) &=& 
   A_4^L(\phi,1_\qb,2_q,3,4) + A_4^R(\phi,1_\qb,2_q,3,4) + A_4^L(\phi,1_\qb,3,2_q,4)  \nonumber\\
&+&A_4^L(\phi,1_\qb,2_q,4,3) + A_4^R(\phi,1_\qb,2_q,4,3) + A_4^L(\phi,1_\qb,4,2_q,3)  \,.
\label{A43defn}
\eea
All of these colour decomposition equations, namely 
Eqs.~(\ref{qqggtreedecomp}, \ref{qqggloopdecomp}, \ref{A41defn}, \ref{A43defn})
are equally valid if the $\phi$ is replaced by a $\phid$ or a Higgs boson $H$. Table ~\ref{Haqggrefs} indicates the references in which the various helicity contributions to $Hgggg$ have first been calculated analytically. The amplitudes for a Higgs boson with four quarks was first calculated analytically in ~\cite{Ellis:2005qe} (with explicit helicity amplitudes in ~\cite{Dixon:2009uk}). 

\TABLE{
\begin{tabular}{|l|l|l|}
\hline
$H$ amplitude & $\phi$ amplitude & $\phid$ amplitude\\
\hline
${\cal A}(H,1_{\bar{q}}^-,2_q^+,3^+,4^+)$
&${\cal A}(\phi,1_{\bar{q}}^-,2_q^+,3^+,4^+)$~\cite{Berger:2006sh}    &${\cal A}(\phid,1_{\bar{q}}^-,2_q^+,3^+,4^+)$~\cite{Badger:2009vh}  \\
${\cal A}(H,1_{\bar{q}}^-,2_q^+,3^-,4^-)$ 
&${\cal A}(\phi,1_{\bar{q}}^-,2_q^+,3^-,4^-)$~\cite{Badger:2009vh} &${\cal A}(\phid,1_{\bar{q}}^-,2_q^+,3^-,4^-)$~\cite{Berger:2006sh}\\
${\cal A}(H,1_{\bar{q}}^-,2_q^+,3^+,4^-)$ 
&${\cal A}(\phi,1_{\bar{q}}^-,2_q^+,3^+,4^-)$~\cite{Dixon:2009uk} &${\cal A}(\phid,1_{\bar{q}}^-,2_q^+,3^+,4^-)$~\cite{Dixon:2009uk}  \\
${\cal A}(H,1_{\bar{q}}^-,2_q^+,3^-,4^+)$ 
&${\cal A}(\phi,1_{\bar{q}}^-,2_q^+,3^-,4^+)$~\cite{Dixon:2009uk} &${\cal A}(\phid,1_{\bar{q}}^-,2_q^+,3^-,4^+)$~\cite{Dixon:2009uk} \\
\hline
\end{tabular}
\caption{$\phi$ and $\phid$ amplitudes needed to construct a given
one-loop $H\bar{q}qgg$ amplitude, together with the references where
they can be obtained.  In all cases the $\phid$ amplitudes are
constructed from the $\phi$ amplitudes given in the reference, using
the parity operation.}\label{Haqggrefs}}

\section{Conclusion}

This report summarises the recent completion of the analytic results for the process $pp\rightarrow H+2j$. The results were obtained by using the unitarity method to calcaulte the cut-constructible pieces of the amplitudes. The rational pieces were calculated with various techniques, for the non-NMHV helicity configurations the unitarity-bootstrap technique was used. For the most recent NMHV calculations the rational pieces were obtained from Feynman diagrams. The calculations were performed using an effective Lagrangian which simplifies the full theory in which the top quark loops are 
included. The effective theory provides an accurate prescription of the physics provided that $m_H < 2m_t$.  It is hoped that the analytic formulae described here will aid in producing fast phenomenological studies of Higgs physics at hadron colliders.

\section*{Acknowledgements}

I would like to thank my collaborators Simon Badger, John Campbell, Keith Ellis, Nigel Glover and Pierpaolo Mastrolia. I would also like to thank Lance Dixon for useful discussions. This work was funded by an STFC studentship.


\begin{thebibliography}{99}



%%\cite{:2009pt}
\bibitem{2009pt}
  [CDF Collaboration and D0 Collaboration],
  %``Combined CDF and DZero Upper Limits on Standard Model Higgs-Boson
  %Production with up to 4.2 fb-1 of Data,''
  arXiv:0903.4001 [hep-ex].
  %%CITATION = ARXIV:0903.4001;%%

%\cite{Djouadi:2005gi}
\bibitem{Djouadi:2005gi}
  {\it For an extensive review of theoretical aspects of Higgs boson production
  and prospects for detection, see:}
  A.~Djouadi,
  %``The Anatomy of electro-weak symmetry breaking. I: The Higgs boson in the
  %standard model,''
  Phys.\ Rept.\  {\bf 457}, 1 (2008)
  [arXiv:hep-ph/0503172]
  and references therein.
  %%CITATION = PRPLC,457,1;%%

%\cite{Ellis:2005qe}
\bibitem{Ellis:2005qe}
  R.~K.~Ellis, W.~T.~Giele and G.~Zanderighi,
  %``Virtual QCD corrections to Higgs boson plus four parton processes,''
  Phys.\ Rev.\  D {\bf 72}, 054018 (2005)
  [Erratum-ibid.\  D {\bf 74}, 079902 (2006)]
  [arXiv:hep-ph/0506196].
  %%CITATION = PHRVA,D72,054018;%%

%\cite{Campbell:2006xx}
\bibitem{Campbell:2006xx}
  J.~M.~Campbell, R.~K.~Ellis and G.~Zanderighi,
  %``Next-to-leading order Higgs + 2 jet production via gluon fusion,''
  JHEP {\bf 0610}, 028 (2006)
  [arXiv:hep-ph/0608194].
  %%CITATION = JHEPA,0610,028;%%

%\cite{Wilczek:1977zn}
\bibitem{Wilczek:1977zn}
  F.~Wilczek,
  %``Decays Of Heavy Vector Mesons Into Higgs Particles,''
  Phys.\ Rev.\ Lett.\  {\bf 39}, 1304 (1977).
  %%CITATION = PRLTA,39,1304;%% 
  


%\cite{Djouadi:1991tka}
\bibitem{Djouadi:1991tka}
  A.~Djouadi, M.~Spira and P.~M.~Zerwas,
  %``Production of Higgs bosons in proton colliders: QCD corrections,''
  Phys.\ Lett.\  B {\bf 264}, 440 (1991).
  %%CITATION = PHLTA,B264,440;%%

%\cite{Dawson:1990zj}
\bibitem{Dawson:1990zj}
  S.~Dawson,
  %``Radiative corrections to Higgs boson production,''
  Nucl.\ Phys.\  B {\bf 359}, 283 (1991).
  %%CITATION = NUPHA,B359,283;%%

%\cite{Dixon:2004za}
\bibitem{Dixon:2004za}
  L.~J.~Dixon, E.~W.~N.~Glover and V.~V.~Khoze,
  %``MHV rules for Higgs plus multi-gluon amplitudes,''
  JHEP {\bf 0412}, 015 (2004)
  [arXiv:hep-th/0411092].
  %%CITATION = JHEPA,0412,015;%%

\bibitem{Badger:HpartonMHV}
S.~D. Badger, E.~W.~N. Glover and V.~V. Khoze, {\it {MHV rules for Higgs plus
  multi-parton amplitudes}},  {\em JHEP} {\bf 03} (2005) 023
  [\href{http://arXiv.org/abs/hep-th/0412275}{{\tt hep-th/0412275}}].
%%CITATION = HEP-TH/0412275;%%


%\cite{Britto:2004nc}
\bibitem{Britto:2004nc}
  R.~Britto, F.~Cachazo and B.~Feng,
  %``Generalized unitarity and one-loop amplitudes in N = 4  super-Yang-Mills,''
  Nucl.\ Phys.\  B {\bf 725}, 275 (2005)
  [arXiv:hep-th/0412103].
  %%CITATION = NUPHA,B725,275;%%

\bibitem{Forde:intcoeffs}
 D.~Forde,
 %{\it {Direct extraction of one-loop integral coefficients}},
 {\em Phys. Rev.} {\bf D75} (2007) 125019
  [\href{http://arXiv.org/abs/0704.1835}{{\tt 0704.1835}}].
%%CITATION = 0704.1835;%%


\bibitem{Britto:ccqcd}
 R.~Britto, B.~Feng and P.~Mastrolia,
 %{\it The cut-constructible part of {QCD} amplitudes},
 {\em Phys. Rev.} {\bf D73} (2006) 105004
  [\href{http://arXiv.org/abs/hep-ph/0602178}{{\tt hep-ph/0602178}}].
%%CITATION = HEP-PH 0602178;%%



\bibitem{Britto:sqcd}
 R.~Britto, E.~Buchbinder, F.~Cachazo and B.~Feng, 
 %{\it One-loop amplitudes of gluons in {SQCD}},
 {\em Phys. Rev.} {\bf D72} (2005) 065012
  [\href{http://arXiv.org/abs/hep-ph/0503132}{{\tt hep-ph/0503132}}].
%%CITATION = HEP-PH 0503132;%%


%\cite{Mastrolia:2009dr}
\bibitem{Mastrolia:2009dr}
  P.~Mastrolia,
  %``Double-Cut of Scattering Amplitudes and Stokes' Theorem,''
  Phys.\ Lett.\  B {\bf 678}, 246 (2009)
  [arXiv:0905.2909 [hep-ph]].
  %%CITATION = PHLTA,B678,246;%%


\bibitem{Berger:genhels}
C.~F. Berger, Z.~Bern, L.~J. Dixon, D.~Forde and D.~A. Kosower, {\it
  Bootstrapping one-loop {QCD} amplitudes with general helicities},  {\em Phys.
  Rev.} {\bf D74} (2006) 036009
  [\href{http://arXiv.org/abs/hep-ph/0604195}{{\tt hep-ph/0604195}}].
%%CITATION = HEP-PH 0604195;%%

\bibitem{Britto:proof}
R.~Britto, F.~Cachazo, B.~Feng and E.~Witten, {\it Direct proof of tree-level
  recursion relation in {Yang-Mills} theory},  {\em Phys. Rev. Lett.} {\bf 94}
  (2005) 181602 [\href{http://arXiv.org/abs/hep-th/0501052}{{\tt
  hep-th/0501052}}].
%%CITATION = HEP-TH 0501052;%%



\bibitem{Dawson:Htomultijet}
S.~Dawson and R.~P. Kauffman,
 %{\it {Higgs} boson plus multi - jet rates at the {SSC}},
 {\em Phys. Rev. Lett.} {\bf 68} (1992) 2273--2276.
%%CITATION = PRLTA,68,2273;%%

%\cite{Bern:1994zx}
\bibitem{Bern:1994zx}
  Z.~Bern, L.~J.~Dixon, D.~C.~Dunbar and D.~A.~Kosower,
  %``One-Loop n-Point Gauge Theory Amplitudes, Unitarity and Collinear Limits,''
  Nucl.\ Phys.\  B {\bf 425} (1994) 217
  [arXiv:hep-ph/9403226].
  %%CITATION = NUPHA,B425,217;%%

%\cite{Badger:2006us}
\bibitem{Badger:2006us}
  S.~D.~Badger and E.~W.~N.~Glover,
  %``One-loop helicity amplitudes for H -> gluons: the all-minus
  %configuration,''
  Nucl.\ Phys.\ Proc.\ Suppl.\  {\bf 160}, 71 (2006)
  [arXiv:hep-ph/0607139].
  %%CITATION = NUPHZ,160,71;%%

%\cite{Badger:2007si}
\bibitem{Badger:2007si}
  S.~D.~Badger, E.~W.~N.~Glover and K.~Risager,
  %``One-loop phi-MHV amplitudes using the unitarity bootstrap,''
  JHEP {\bf 0707}, 066 (2007)
  [arXiv:0704.3914 [hep-ph]].
  %%CITATION = JHEPA,0707,066;%%

%\cite{Glover:2008ffa}
\bibitem{Glover:2008ffa}
  E.~W.~N.~Glover, P.~Mastrolia and C.~Williams,
  %``One-loop phi-MHV amplitudes using the unitarity bootstrap: the general
  %helicity case,''
  JHEP {\bf 0808}, 017 (2008)
  [arXiv:0804.4149 [hep-ph]].
  %%CITATION = JHEPA,0808,017;%%

%\cite{Berger:2006sh}
\bibitem{Berger:2006sh}
  C.~F.~Berger, V.~Del Duca and L.~J.~Dixon,
  %``Recursive construction of Higgs+multiparton loop amplitudes: The last of
  %the phi-nite loop amplitudes,''
  Phys.\ Rev.\  D {\bf 74}, 094021 (2006)
  [Erratum-ibid.\  D {\bf 76}, 099901 (2007)]
  [arXiv:hep-ph/0608180].
  %%CITATION = PHRVA,D74,094021;%%

%\cite{Badger:2009hw}
\bibitem{Badger:2009hw}
  S.~Badger, E.~W.~N.~Glover, P.~Mastrolia and C.~Williams,
  %``One-loop Higgs plus four gluon amplitudes: Full analytic results,''
   JHEP {\bf 1001}, 036 (2010)
  arXiv:0909.4475 [hep-ph].
  %%CITATION = ARXIV:0909.4475;%%

%\cite{Bern:1994fz}
\bibitem{Bern:1994fz}
  Z.~Bern, L.~J.~Dixon and D.~A.~Kosower,
  %``One Loop Corrections To Two Quark Three Gluon Amplitudes,''
  Nucl.\ Phys.\  B {\bf 437}, 259 (1995)
  [arXiv:hep-ph/9409393].
  %%CITATION = NUPHA,B437,259;%%
  
%\cite{Dixon:2009uk}
\bibitem{Dixon:2009uk}
  L.~J.~Dixon and Y.~Sofianatos,
  %``Analytic one-loop amplitudes for a Higgs boson plus four partons,''
   JHEP {\bf 0908}, 058 (2009)
  arXiv:0906.0008 [hep-ph].
  %%CITATION = ARXIV:0906.0008;%%


%\cite{Badger:2009vh}
\bibitem{Badger:2009vh}
    S.~Badger, J.~M.~Campbell, R.~K.~Ellis and C.~Williams,
            % "Analytic results for the one-loop NMHV Hqqgg amplitude",
   JHEP {\bf 0912}, 035 (2009)
  [arXiv:0910.4481 [hep-ph]].    
     %%CITATION = 0910.4481;%%"
\end{thebibliography}
\end{document}